# Theory of Confined States of Positronium in Spherical and Circular Quantum Dots with Kane's Dispersion Law

K.G. Dvoyan, S.G. Matinyan, B. Vlahovic

Department of Physics, North Carolina Central University, 1801 Fayetteville St., Durham, NC, 27707, USA

**Abstract**

Confined states of a positronium (Ps) in the spherical and circular quantum dots (QDs) are theoretically investigated in two size-quantization regimes: strong and weak. Two-band approximation of Kane dispersion law and parabolic dispersion law of charge carriers are considered. It is shown that the electron-positron pair instability is a consequence of dimensionality reduction, not of the size quantization (SQ). The binding energies for the Ps in circular and spherical QDs are calculated. The criterion of the Ps formation at a given radius of a QD is derived.

## 1. Introduction.

Investigation of new physical properties of zero-dimensional objects, particularly semiconductor quantum dots (QDs), is a fundamental part of modern physics. Extraordinary properties of nanostructures are mainly a consequence of quantum confinement effects. A lot of theoretical and experimental works are devoted to the study of the electronic, impurity, excitonic and optical properties of semiconductor QDs. Potential applications of various nanostructures in optoelectronic and photonic devices are predicted and under intensive study of many research groups [1-7]. In low-dimensional structures along with the size quantization (SQ) effects one often deals with the Coulomb interaction between the charge carriers (CC). SQ can successfully compete with Coulomb quantization and even prevails over it in certain cases. In the Coulomb problems in the SQ systems one has to use different quantum mechanical approaches along with numerical methods. Thus, the significant difference between the effective masses of the impurity (holes) and electron allows us to use the Born-Oppenheimer approximation [8,9]. When the energy conditioned by the SQ is much more the Coulomb energy, the problem is solved in the framework of perturbation theory, where the role of a small correction plays the term of the Coulomb interaction in the problem Hamiltonian [10].

The situation is radically changed when the effective mass of the impurity center (hole) is comparable to the mass of the electron. For example, in the narrow-gap semiconductors for which the CC standard (parabolic) dispersion low is violated, the effective masses of the electron and light hole are equal [11-14]. It is obvious that in the case of equal effective masses adiabatic approximation is not applicable. A similar situation arises in considering the Coulomb interaction of the electron-positron pair. Experimental detection of anti-particles – particles with negative energies is one of the most important scientific discoveries of the 20[th] century [15]. Obviously, the anti-particles doping in semiconductor systems with reduced dimensionality greatly increases the possibilities of the external manipulation of the physical properties of these nanostructures and many times broaden the area of potential applications of devices based on them. On the other hand, such an approach makes a real the study of the changes of the anti-particles properties and properties of their complexes formed in semiconductor media under the influence of SQ. It is known that the antiparticles and particles have opposite additive quantum numbers (e.g. electric charge, lepton number etc), but are otherwise identical. Combinations of particle-antiparticle pairs may form exotic atomic states, the most well-known example being positronium (Ps), the bound state between an electron and positron [16]. There are two types of Ps: ortho- (for the parallel orientation of the spins) and para- (if antiparallel orientation). Orthopositronium has a lifetime $\tau \sim 1.4 \cdot 10^{-7} s$ and annihilates with the emission of three gamma quanta, which is by three orders exceeds the lifetime of parapositronium [17-19]. Although Ps will decay via self-annihilation, its lifetime is long enough that it has a well defined atomic structure. Thus, in [20-23] the authors experimentally detected the occurrence of a positronium and its molecules in the structure of porous silicon and also detected positron lines of light absorption. Wheeler supposed that two positronium atoms might combine to form the dipositronium molecule [24]. Schrader theoretically studied this molecule [25]. Because Ps has a short lifetime and it is difficult to obtain low energy positrons in large numbers, dipositronium has not been observed unambiguously. Mills and Cassidy's group showed that dipositronium was created on the internal pore surfaces when intense positron bursts are implanted into a thin film of porous silica. Moreover, in Ref [26] authors report observations of transitions between the ground state of $Ps_2$ and excited state. These results experimentally confirm the existence of the dipositronium molecule. As a purely leptonic, macroscopic quantum matter–antimatter system, this would be of interest in its own right, but it would also represent a milestone on the path to produce an annihilation gamma-ray laser [27]. Further, in the work [21] porous silica film contains interconnected pores with a diameter $d < 4$ nm. From above mentioned follows, that it is logically necessary to discuss size quantization effects related with this topic. In Ref [28] the additional quantization effects on the Ps states conditioned by the QD confinement have been revealed along with quantization conditioned by the Coulomb interaction in the framework of the standard (parabolic) dispersion low of CCs.

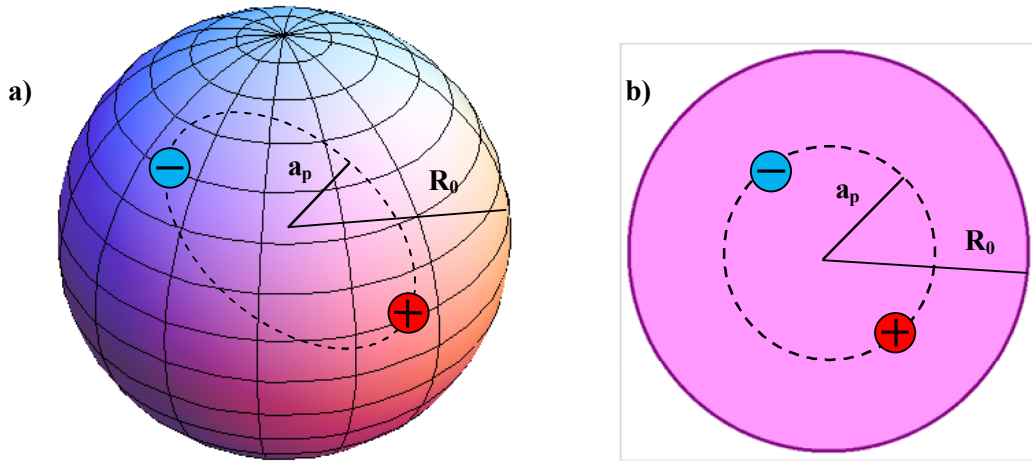

Fig. 1 The electron-positron pair: a) in the spherical QD, b) in the circular QD.

In [29], the authors first observed experimentally the Bloch states of positronium in alkali halides and measured its effective mass in *NaBr* and *RbCl* crystals. In particular the temperature dependence of the transition from a self-trapped Ps to the Bloch state is investigated. It is natural to assume that creating a leap of the potentials on the boundaries of the media with the selection of specific materials with different widths of the band gaps it will be possible to localize Bloch state of the Ps in a variety of nanostructures. There are many works devoted to the study of the Ps states in various solids or on their surfaces. For example, in [30] the work functions of the positron and positronium for metals and semiconductors are calculated. Calculations of positron energy levels and work functions of the positron and positronium in the case of narrow gap semiconductors are given in the paper [31]. It should be noted that in the narrow-gap semiconductors, in addition to reduction of the band gap, the dispersion law of CCs is complicated as well. However, there are quite a number of papers in which more complicated dependence of the CC effective mass on the energy is considered [11-14, 32-35] in the framework of the Kane's theory. For example, for the narrow-gap QDs of *InSb* the dispersion law of CCs is nonparabolic and it is well described by the Kane two-band mirror model [14, 36]. Within the framework of the two-band approximation the electron (light hole) dispersion law formally coincides with the relativistic law. It is known that in the case of the Kane dispersion law the binding energy of the impurity center turns out more than in the case of a parabolic law [36, 37]. It is also known that the reduction of the system dimensionality leads to increase in the Coulomb quantization. Hence, in the 2D case the ground state binding energy of the impurity increases four times compared to that of the 3D case [38].

As the foregoing theoretical analysis of Ps shows, the investigation of quantum states in the SQ semiconductor systems with Kane's dispersion low is a prospective problem of the modern nanoscience. In the present paper the quantum states of the electron-positron pair in the spherical and circular QDs consisting of *InSb* and *GaAs* with impermeable walls are considered. The quantized states of both a Ps and individually quantized electron and positron are discussed in the two SQ regimes – weak and strong, respectively.

## 2. Theory.

### 2.1. Positronium in a spherical QD with Kane's dispersion law.

Let us consider an impermeable spherical QD. The potential energy of a particle in the spherical coordinates has the following form:

$$U(\rho,\theta,\varphi) = \begin{cases} 0, \rho \leq R_0 \\ \infty, \rho > R_0 \end{cases}, \quad (1)$$

where $R_0$ is the radius of a QD. The radius of a QD and effective Bohr radius of a Ps $a_p$ play the role of the problem parameters, which radically affect the behavior of the particle inside a QD. In our model, the criterion of a Ps formation possibility is the ratio of the Ps effective Bohr radius and QD radius. In what follows, we analyze the problem in two size quantization (SQ) regimes: strong and weak.

### 2.1.1 Strong size quantization regime.

In the regime of strong SQ, when the condition $R_0 \ll a_p$ takes place, the energy of the Coulomb interaction between an electron and positron is much less than the energy caused by the SQ contribution. In this approximation the Coulomb interaction between the electron and positron can be neglected. Then the problem reduces to the determination of an electron and positron energy states separately. As noted above, the dispersion low for narrow-gap semiconductors is nonparabolic and is given in the following form [11, 32]:

$$E^2 = P^2 S^2 + m_{e(p)}^{*2} S^4, \qquad (2)$$

where $S \sim 10^8 \, sm/sec$ is the parameter related with the semiconductor band gap $E_g = 2m_e^* S^2$. Let us write the Klein-Gordon equation [39] for a spherical QD consisting of $InSb$ with electron and positron when their Coulomb interaction is neglected:

$$\left( \left( P_e^2 + P_p^2 \right) S^2 + \left( m_e^{*2} + m_p^{*2} \right) S^4 \right) \Psi \left( \vec{r}_e, \vec{r}_p \right) = E^2 \Psi \left( \vec{r}_e, \vec{r}_p \right), \qquad (3)$$

where $P_{e(p)}$ is the momentum operator of the particle (electron, positron), $m_{e(p)}^*$ is the effective mass of the particle, $E$ is total energy of the system. After simple transformations the equation (3) can be written as the reduced Schrödinger equation:

$$\left( -\frac{1}{2} \nabla_e^2 - \frac{1}{2} \nabla_p^2 \right) \Psi \left( \vec{r}_e, \vec{r}_p \right) = \varepsilon_0 \Psi \left( \vec{r}_e, \vec{r}_p \right), \qquad (4)$$

where the $\varepsilon_0 = \dfrac{2\varepsilon^2 - \varepsilon_g^2}{2\varepsilon_g}$, $\varepsilon = \dfrac{E}{E_R^p}$ notations are introduced. The wave function of the problem is sought in the form $\Psi \left( \vec{r}_e, \vec{r}_p \right) = \Psi_e \left( \vec{r}_e \right) \Psi_p \left( \vec{r}_p \right)$. After separation of variables, one can obtain the following equation for the electron:

$$\left( \nabla_e^2 + 2\varepsilon_e \right) \Psi_e \left( \vec{r}_e \right) = 0. \qquad (5)$$

Seeking the wave function in the form $\Psi_e \left( \vec{r}_e \right) = R(r) Y_{lm} (\theta, \varphi)$, the following equation for the radial part of (5) could be obtained:

$$R''(r) + \frac{2}{r} R'(r) + \left( 2\varepsilon_e - \frac{l(l+1)}{r^2} \right) R(r) = 0. \qquad (6)$$

Here $r = \frac{\rho}{a_p}$, $l$ is orbital quantum number, $m$ is magnetic quantum number, $\mu = \frac{m_e^* m_p^*}{m_e^* + m_p^*}$ is the reduced mass of a Ps, $E_R^p = \frac{\hbar^2}{2\mu a_p^2} = \frac{\hbar^2}{m_e^* a_p^2} = \frac{e^2}{2\kappa a_p}$ is the effective Rydberg energy of a Ps, $\kappa$ is the dielectric constant of the semiconductor, $a_p = \frac{\kappa \hbar^2}{\mu e^2} = \frac{2\kappa \hbar^2}{m_p^* e^2}$ is a Ps effective Bohr radius, $\varepsilon_e = \frac{E_e}{E_R^p}$ is dimensionless energy, $\varepsilon_g = \frac{E_g}{E_R^p} = \frac{2a_p^2}{\lambdabar_C^2} = \frac{8}{\alpha_0^2}$ is dimensionless band gap width, $\alpha_0 = \frac{e^2}{\hbar S \kappa}$ is the analogue of fine structure constant and $\lambdabar_C = \frac{\hbar}{m_e^* S}$ is the analogue of Compton wavelength in a narrow band gap semiconductors with Kane's dispersion law. Solving the equation (6), taking into account the boundary conditions one can obtain for the wave functions:

$$\Psi_e(\vec{r}_e) = \frac{1}{\sqrt{2\pi} r_0 J_{l+3/2}(\sqrt{2\varepsilon_e} r_0)} \frac{J_{l+1/2}(\sqrt{2\varepsilon_e} r)}{\sqrt{r}} Y_{lm}(\theta, \varphi), \qquad (7)$$

where $r_0 = \frac{R_0}{a_p}$, $J_{l+1/2}(z)$ are Bessel functions of half-integer arguments, $Y_{lm}(\theta, \varphi)$ are spherical functions [40]. The following result could be revealed for the electron eigenvalues:

$$\varepsilon_e = \frac{\alpha_{n,l}^2}{2 r_0^2}, \qquad (8)$$

where $\alpha_{n,l}$ are the roots of the Bessel functions. The electron energy (8) is a constant of separation of variables in the positron reduced Schrödinger equation:

$$\left(\nabla_p^2 + 2(\varepsilon_0 - \varepsilon_e)\right)\Psi_p(\vec{r}_p) = 0. \qquad (9)$$

Solving the equation in a similar way, finally, in the strong SQ regime one can derive the following expression for the total energy of the particles' system:

$$\varepsilon = \varepsilon_g \sqrt{\frac{\alpha_{n,l}^2 + \alpha_{n',l'}^2}{r_0^2} \frac{1}{2\varepsilon_g} + \frac{1}{2}}. \qquad (10)$$

Here $n,l$ ($n',l'$) are the main and orbital quantum numbers, correspondingly. In more convenient units $\varepsilon_g$ and $\lambdabar_C$, the expression of energy (10) can be written in a simpler form suitable for graphical representations:

$$\tilde{\varepsilon} = \sqrt{\frac{\alpha_{n,l}^2 + \alpha_{n',l'}^2}{4\tilde{r}_0^2} + \frac{1}{2}}, \qquad (11)$$

where $\tilde{\varepsilon} = \frac{E}{E_g}$, $\tilde{r}_0 = \frac{R_0}{\lambdabar_C}$. For comparison (see (10)), in the case of a parabolic dispersion law (for QD consisting of $GaAs$) the total energy in the strong SQ is given as [28]:

$$\varepsilon_{Par} = \frac{\alpha_{n,l}^2 + \alpha_{n',l'}^2}{2r_0^2}. \qquad (12)$$

**2.1.2. Weak size quantization regime.**

In this regime, when the condition $R_0 \gg a_p$ takes place, the system's energy is caused mainly by the electron-positron Coulomb interaction. In other words, we consider the motion of a Ps as a whole in a QD. In the case of the presence of Coulomb interaction between an electron and positron, the Klein-Gordon equation can be written as [37]

$$\left((P_e^2 + P_p^2)S^2 + (m_e^{*2} + m_p^{*2})S^4\right)\Psi(\vec{r}_e, \vec{r}_p) = \left(E + \frac{e^2}{\kappa|\vec{r}_e - \vec{r}_p|}\right)^2 \Psi(\vec{r}_e, \vec{r}_p), \qquad (13)$$

where $e$ is the elementary charge. After simple transformations, as in the case of a strong SQ regime, the Klein-Gordon equation reduces to the Schrödinger equation with a certain effective energy, and then the wave function of the system can be represented as

$$\Psi(\vec{r}_e, \vec{r}_p) = \psi(\vec{r})\Phi(\vec{R}), \qquad (14)$$

where $\vec{r} = \vec{r}_e - \vec{r}_p$, $\vec{R} = \dfrac{m_e^* \vec{r}_e + m_p^* \vec{r}_p}{m_e^* + m_p^*}$. Here $\psi(\vec{r})$ describes the relative motion of the electron and positron, while $\Phi(\vec{R})$ describes the motion of the Ps center of gravity. After switching to the new coordinates, the Schrödinger equation takes the following form:

$$\left(-\frac{\hbar^2}{2M_0}\nabla_{\vec{R}}^2 - \frac{\hbar^2}{2\mu}\nabla_{\vec{r}}^2\right)\psi(\vec{r})\Phi(\vec{R}) = \left[\frac{\left(E + \dfrac{e^2}{\kappa|\vec{r}|}\right)^2 - \left(m_e^{*2} + m_p^{*2}\right)S^4}{2m_e^* S^2}\right]\psi(\vec{r})\Phi(\vec{R}), \quad (15)$$

where $M_0 = m_e^* + m_p^*$ is the mass of a Ps. One can derive the equation for a Ps center of gravity, after separation of variables, in the $E_R^p$ and $a_p$ units:

$$-\frac{1}{4}\nabla_{\vec{R}}^2 \Phi(\vec{R}) = \varepsilon_R \Phi(\vec{R}), \quad (16)$$

or

$$\Phi''(R) + \frac{2}{R}\Phi'(R) + \left(4\varepsilon_R - \frac{L(L+1)}{R^2}\right)\Phi(R) = 0, \quad (17)$$

where $\varepsilon_R$ is the energy of a Ps center of gravity quantized motion, and $L$ is the orbital quantum number of a Ps motion as a whole. For energy and wave functions of the electron-positron pair center of gravity motion one can obtain, respectively, the following expressions:

$$\varepsilon_R = \frac{\alpha_{N,L}^2}{4r_0^2}, \quad (18)$$

$$\Phi(\vec{R}) = \frac{1}{\sqrt{2\pi r_0} J_{L+3/2}\left(\sqrt{4\varepsilon_R} r_0\right)} \frac{J_{L+1/2}\left(\sqrt{4\varepsilon_R} R\right)}{\sqrt{R}} Y_{LM}(\theta, \varphi), \quad (19)$$

where $N$ and $M$ are, respectively, the main and magnetic quantum numbers of a Ps motion as a whole.

Further, let us consider the relative motion of the electron-positron pair. The wave function of the problem is sought in the form $\psi(\vec{r}) = \frac{1}{\sqrt{r}}\chi(r)Y_{lm}(\theta,\varphi)$. After simple transformations, the radial part of the reduced Schrödinger equation can be written as:

$$\chi''(r) + \frac{1}{r}\chi'(r) + \left(\varepsilon' - \frac{\left(l+\frac{1}{2}\right)^2 - \beta}{r^2} + \frac{\alpha}{r}\right)\chi(r) = 0, \qquad (20)$$

where the following notations are introduced: $\varepsilon' = \frac{2\varepsilon^2 - \varepsilon_g^2}{2\varepsilon_g} - \varepsilon_R$, $\alpha = \frac{4\varepsilon}{\varepsilon_g}$, $\beta = \frac{4}{\varepsilon_g}$, $\varepsilon = \frac{E}{E_R^p}$. The change of variable $\xi = 2\sqrt{-\varepsilon'}r$ transforms the equation (20) to

$$\chi''(\xi) + \frac{1}{\xi}\chi'(\xi) + \left(-\frac{1}{4} - \frac{\left(l+\frac{1}{2}\right)^2 - \beta}{\xi^2} + \frac{\gamma}{\xi}\right)\chi(\xi) = 0, \qquad (21)$$

where the parameter $\gamma = \frac{\alpha}{2\sqrt{-\varepsilon'}}$ is introduced. When $\xi \to 0$ the desired solution of (21) is sought in the form $\chi(\xi \to 0) = \chi_0 \sim \xi^\lambda$ [40,41]. Substituting this in equation (21) one get a quadratic equation with two solutions:

$$\lambda_1 = -\sqrt{\left(l+\frac{1}{2}\right)^2 - \beta}, \quad \lambda_2 = \sqrt{\left(l+\frac{1}{2}\right)^2 - \beta}. \qquad (22)$$

The solution satisfying the finiteness condition of the wave function is given as $\chi_0 \sim \xi^{\sqrt{\left(l+\frac{1}{2}\right)^2 - \beta}}$. When $\xi \to \infty$ the equation (21) takes the form: $\chi''(\xi) - \frac{1}{4}\chi(\xi) = 0$. The solution satisfying the standard conditions can be written as $\chi(\xi \to \infty) = \chi_\infty \sim e^{-\xi/2}$ [42]. Thus, the solution is sought in the form:

$$\chi(\xi) = \xi^{\lambda} e^{-\xi/2} f(\xi). \tag{23}$$

Substituting the function (23) in equation (21) one can get the Kummer equation [40]:

$$\xi f''(\xi) + (2\lambda + 1 - \xi) f'(\xi) + \left(\gamma - \lambda - \frac{1}{2}\right) f(\xi) = 0, \tag{24}$$

which solutions are given by the first kind degenerate hypergeometric functions:

$$f(\xi) = {}_1F_1\left(-\left(\gamma - \lambda - \frac{1}{2}\right), 2\lambda + 1, \xi\right). \tag{25}$$

The expression $\gamma - \lambda - \frac{1}{2}$ needs to be a non-negative integer $n_r$ (radial quantum number) providing the finiteness of the wave functions everywhere:

$$n_r = \gamma - \lambda - \frac{1}{2}, \quad n_r = 0, 1, 2, \dots. \tag{26}$$

From the condition (26) for the positron energy in a *InSb* spherical QD with a non-parabolic dispersion low (in dimensionless units) one can derive the following expression:

$$\varepsilon_{Ps}^{Kane} = -\frac{\sqrt{\varepsilon_g + \frac{\alpha_{N,L}^2}{2r_0^2}}}{\sqrt{\frac{2}{\varepsilon_g} + \frac{8}{\varepsilon_g^2\left(n_r + \sqrt{\left(l + \frac{1}{2}\right)^2 - \frac{4}{\varepsilon_g} + \frac{1}{2}}\right)^2}}}. \tag{27}$$

The expression of a Ps energy in a spherical QD with a parabolic dispersion law obtained in the work [28] is given for comparison:

$$\varepsilon_{Ps}^{Par} = \frac{\alpha_{n,l}^2}{4r_0^2} - \frac{1}{N'^2}, \tag{28}$$

where $N'$ is the main quantum number of electron-positron pair relative motion under influence of Coulomb interaction only.

Determining the binding energy as the energy difference between the cases of the presence and absence of positron in a QD, one obtains finally the following expression:

$$\varepsilon_{Bind}^{Kane} = \varepsilon_g \sqrt{\frac{\alpha_{n',l'}^2}{r_0^2}\frac{1}{2\varepsilon_g} + \frac{1}{2}} + \frac{\sqrt{\varepsilon_g + \frac{\alpha_{N,L}^2}{2r_0^2}}}{\sqrt{\frac{2}{\varepsilon_g} + \frac{8}{\varepsilon_g^2\left(n_r + \sqrt{\left(l+\frac{1}{2}\right)^2 - \frac{4}{\varepsilon_g}} + \frac{1}{2}\right)^2}}}. \quad (29)$$

For clarity, it makes sense to compare this expression to similar result obtained in the case of a parabolic dispersion law [28]:

$$\varepsilon_{Bind}^{Par} = \frac{\alpha_{n,l}^2}{4r_0^2} + \frac{1}{N'^2}. \quad (30)$$

Here it is necessary to note some important remarks in contrast to the case of the problem of hydrogen-like impurities in a semiconductor with Kane's dispersion law, considered in [42,45], in the case of 3D positron, the instability of the ground state energy is absent. Thus, in the case of hydrogen-like impurity, the electron energy becomes unstable when $Z\alpha_0 > \frac{1}{2}$ ($Z$ is a charge number), and the phenomenon of the particle falling into center takes place. However, in our case the expression $\left(l+\frac{1}{2}\right)^2 - \frac{4}{\varepsilon_g}$ under the square root (see (27)) does not become negative even for the ground state with $l = 0$. In other words, in the case of a 3D Ps with Kane's dispersion low, it would be necessary a fulfillment of condition for the analogue of fine structure constant $\alpha_0 > \frac{1}{\sqrt{2}}$ to obtain instability in the ground state. However, obviously it is impossible for the QD consisting of $InSb$, for which the analogue of fine structure constant is $\alpha_0 = 0.123$. It should be noted also that the instability is absent even at the temperature $T = 300K$, when the band gap width is less and equals to $E_g = 0.17\,eV$ instead of $0.23\,eV$, which is realized at the lower temperatures.

Second, for the *InSb* QD, the energy of SQ motion of a Ps center of gravity enters the expression of the energy (binding energy) under the square root, whereas in the parabolic dispersion low case, this energy appears as a simple sum (see (27) and (28) or (29) and (30)).

Third, the Ps energy depends only on the main quantum number of the Coulomb motion in the case of the parabolic dispersion, whereas in the case of Kane's dispersion low it reveals a rather complicated dependence on the radial and orbital quantum numbers. In other words, the non-parabolicity account of the dispersion leads to the removal of "accidental" Coulomb degeneracy in the orbital quantum number [44], however, the energy degeneracy remains in the magnetic quantum number in both cases as a consequence of the spherical symmetry.

For a more detailed analysis of the influence of QD walls on the Ps motion, also consider the case of the "free" Ps in the bulk semiconductor with Kane's dispersion low.

### 2.1.3. A "Free" positronium regime (positronium in a bulk semiconductor).

Klein-Gordon equation for a free atom of Ps can be written as (13). After separating the variables, the wave function is sought in the form (14), where $\vec{r} = \vec{r}_e - \vec{r}_p$, $\vec{R} = \dfrac{m_e^* \vec{r}_e + m_p^* \vec{r}_p}{m_e^* + m_p^*} = 0$. Here again $\psi(\vec{r})$ describes the relative motion of the electron and positron, while $\Phi(\vec{R}) \sim e^{iKR}$ describes the "free" motion of a Ps center of gravity. Similar to (20) after simple transformations one can obtain

$$\chi''(r) + \frac{2}{r}\chi'(r) + \left(\varepsilon_0 - \frac{l(l+1)-\beta}{r^2} + \frac{\alpha}{r}\right)\chi(r) = 0. \qquad (31)$$

Repeating the calculations described above, one can derive the expression for the wave functions:

$$f(\xi) = {}_1F_1\left(-(\gamma - \lambda' - 1), 2\lambda' + 2, \xi\right), \qquad (32)$$

where $\lambda' = \frac{1}{2}\left(-1 + \sqrt{(2l+1)^2 - \frac{16}{\varepsilon_g}}\right)$. The energy of a "free" Ps atom in a narrow bind gap semiconductor with Kane's dispersion low can be revealed from standard conditions:

$$\varepsilon_{Ps}^{Free} = -\frac{\varepsilon_g}{\sqrt{2 + \dfrac{8}{\left(\varepsilon_g\left(n_r + \sqrt{\left(l+\dfrac{1}{2}\right)^2 - \dfrac{4}{\varepsilon_g}} + \dfrac{1}{2}\right)\right)^2}}}. \tag{33}$$

As expected, the expression (33) follows from (27) in the limit case $r_0 \to \infty$. For a clearer identification of the contribution of the SQ in a Ps energy, let us define the confinement energy as a difference between absolute values of energies of a Ps in a spherical QD and a "free" Ps:

$$\varepsilon_{Ps}^{Conf} = \frac{\sqrt{\varepsilon_g\left(\varepsilon_g + \dfrac{\alpha_{N,L}^2}{2r_0^2}\right)} - \varepsilon_g}{\sqrt{2 + \dfrac{8}{\left(\varepsilon_g\left(n_r + \sqrt{\left(l+\dfrac{1}{2}\right)^2 - \dfrac{4}{\varepsilon_g}} + \dfrac{1}{2}\right)\right)^2}}}. \tag{34}$$

It is follows from (34) that in the limiting case $r_0 \to \infty$, the confinement energy becomes zero, as expected. However, it becomes significant in the case of small radius of QD. Note also that the confinement energy defined here should not be confused with the binding energy of a Ps, since the latter, unlike the first, in the limiting case does not become zero.

**2.2. Positronium in two dimensional QD.**

As noted above, the dimensionality reduction dramatically changes the energy of charged particles. Thus, the Coulomb interaction between the impurity center and the electron increases significantly (up to 4 times in the ground state) [38]. Therefore it is interesting to consider the influence of the SQ in the case of 2D interaction of the electron and positron with the non-parabolic dispersion low.

Consider an electron-positron pair in an impermeable 2D circular QD with a radius $R_0$. The potential energy is written as

$$U(\rho,\varphi) = \begin{cases} 0, \rho \leq R_0 \\ \infty, \rho > R_0 \end{cases}. \tag{35}$$

The radius of QD and effective Bohr radius of the Ps $a_p$ again play the role of the problem parameters, which radically affect the behavior of the particle inside a 2D QD.

**2.2.1. Strong size quantization regime.**

As it mentioned, the Coulomb interaction between the electron and positron can be neglected in this approximation. The situation is similar to the 3D case, with the only difference being that the Bessel equation is obtained for radial part of the reduced Schrödinger equation:

$$R''(\eta) + \frac{1}{\eta}R'(\eta) + \left(1 - \frac{m^2}{\eta^2}\right)R(\eta) = 0, \tag{36}$$

and solutions are given by the Bessel functions of the first kind $J_m(\eta)$, where $\eta = \sqrt{2\varepsilon_e}r$. For the electron energy the following expression is obtained:

$$\varepsilon_{e\,2D} = \frac{\alpha_{n_r,m}^2}{2r_0^2}, \tag{37}$$

where $\alpha_{n_r,m}$ are zeroes of the Bessel functions of the integer argument. The following result can be derived for the system total energy:

$$\varepsilon_{2D}^{Kane} = \varepsilon_g \sqrt{\frac{\alpha_{n_r,m}^2 + \alpha_{n_r',m'}^2}{r_0^2}\frac{1}{2\varepsilon_g} + \frac{1}{2}}. \tag{38}$$

Here $n_r, m$ ($n_r', m'$) are the radial and magnetic quantum numbers, respectively. For comparison, in the case of parabolic dispersion law for the 2D pair in a circular QD in the strong SQ regime one can get:

$$\varepsilon_{2D}^{Par} = \frac{\alpha_{n_r,m}^2 + \alpha_{n_r',m'}^2}{2r_0^2}. \qquad (39)$$

**2.2.2. Weak size quantization regime.**

In this case, again the system's energy is caused mainly by the electron-positron Coulomb interaction, and we consider the motion of a Ps as a whole in a QD. Solving the Klein-Gordon equation for this case the wave functions of the system again can be represented in the form (14), however, it must be taken into account that as $\vec{r}_e$ and $\vec{r}_p$, also $\vec{r}$ and $\vec{R}$ are 2D vectors now. Then, the wave functions and a Ps energy of the center of gravity motion, respectively, can be obtained in the 2D case:

$$\Phi(\vec{R}) = \frac{1}{\sqrt{\pi} r_0 J_{M+1}\left(2\sqrt{\varepsilon_{R_{2D}}} r_0\right)} J_M\left(2\sqrt{\varepsilon_{R_{2D}}} R\right) e^{iM\varphi}, \qquad (40)$$

$$\varepsilon_{R_{2D}} = \frac{\alpha_{N,M}^2}{4r_0^2}. \qquad (41)$$

Next, consider the relative motion of the electron-positron pair. Seeking the wave functions of the problem in the form $\psi(\vec{r}) = e^{im\varphi} \chi(r)$, after some transformations the radial part of the reduced Schrodinger equation can be written as:

$$\chi''(\xi) + \frac{1}{\xi}\chi'(\xi) + \left(-\frac{1}{4} - \frac{m^2 - \beta}{\xi^2} + \frac{\gamma}{\xi}\right)\chi(\xi) = 0. \qquad (42)$$

At $\xi \to 0$ the solution of (42) sought in the form $\chi(\xi \to 0) = \chi_0 \sim \xi^\lambda$. Here, in contrast to the equation (21) the quadratic equation is obtained with the following solutions:

$$\lambda_1 = -\sqrt{m^2 - \beta}, \quad \lambda_2 = \sqrt{m^2 - \beta}. \qquad (43)$$

In 2D case, the solution satisfying the condition of finiteness of the wave function is given as $\chi_0 \sim \xi^{\sqrt{m^2-\beta}}$. At $\xi \to \infty$, proceeding analogously to the solution of the equation (21), one should again arrive at the equation of Kummer (24), but with different parameter $\lambda$. Finally, for the energy of the 2D Ps with the Kane dispersion low one can get:

$$\varepsilon_{Ps\,2D}^{Kane} = -\frac{\sqrt{\varepsilon_g + \frac{\alpha_{N,M}^2}{2r_0^2}}}{\sqrt{\frac{2}{\varepsilon_g} + \frac{8}{\varepsilon_g^2\left(n_r + \sqrt{m^2 - \frac{4}{\varepsilon_g}} + \frac{1}{2}\right)^2}}}. \tag{44}$$

A similar result for the case of a parabolic dispersion law is written as:

$$\varepsilon_{Ps\,2D}^{Par} = \frac{\alpha_{N,M}^2}{4r_0^2} - \frac{1}{\left(n_r + |m| + \frac{1}{2}\right)^2} = \frac{\alpha_{N,M}^2}{4r_0^2} - \frac{1}{\left(N' + \frac{1}{2}\right)^2}. \tag{45}$$

Again, determining the binding energy as the energy difference between cases of presence and absence of positron in a QD, one finally obtains the expression:

$$\varepsilon_{Bind\,2D}^{Kane} = \varepsilon_g \sqrt{\frac{\alpha_{n'_r,m'}^2}{r_0^2}\frac{1}{2\varepsilon_g} + \frac{1}{2}} + \frac{\sqrt{\varepsilon_g + \frac{\alpha_{N_r,M}^2}{2r_0^2}}}{\sqrt{\frac{2}{\varepsilon_g} + \frac{8}{\varepsilon_g^2\left(n_r + \sqrt{m^2 - \frac{4}{\varepsilon_g}} + \frac{1}{2}\right)^2}}}. \tag{46}$$

In the case of "free" 2D Ps with Kane dispersion low the energy is:

$$\varepsilon_{Ps\,2D}^{Free} = -\frac{\varepsilon_g}{\sqrt{2 + \frac{8}{\varepsilon_g\left(n_r + \sqrt{m^2 - \frac{4}{\varepsilon_g}} + \frac{1}{2}\right)^2}}}. \tag{47}$$

Here again, the expression (47) follows from (44) at the limit $r_0 \to \infty$. Define again the confinement energy in 2D case as the difference between the absolute values of the Ps energy in a circular QD and a "free" Ps energy:

$$\varepsilon_{Ps_{2D}}^{Conf} = \frac{\sqrt{\varepsilon_g\left(\varepsilon_g + \frac{\alpha_{N,M}^2}{2r_0^2}\right)} - \varepsilon_g}{\sqrt{2 + \frac{8}{\varepsilon_g\left(n_r + \sqrt{m^2 - \frac{4}{\varepsilon_g} + \frac{1}{2}}\right)^2}}}. \qquad (48)$$

Here is also necessary to note two important remarks. First, in contrast to the 3D Ps case, all states with $m = 0$ are unstable in a semiconductor with Kane dispersion law. It is also important that instability is the consequence not only of the dimension reduction of the sample, but also the change of the dispersion law. In other words, "the particle falling into center" [41], or more correctly, the annihilation of the pair in the states with $m = 0$ is the consequence of interaction of energy bands. Thus, the dimension reduction leads to the fourfold increase in the Ps ground state energy in case of parabolic dispersion law, but in the case of Kane dispersion low, annihilation is also possible. Note also that the presence of SQ does not affect the occurrence of instability as it exists both in presence and in absence (see (44) and (47)).

Second, the account of the bands interaction removes the degeneracy of the magnetic quantum number. However, the twofold degeneracy of $m$ of energy remains. Thus, in the case of Kane dispersion law the Ps energy depends on $m^2$, whereas in the parabolic case it depends on $|m|$. Due to the circular symmetry of the problem, the twofold degeneracy of energy remains in both cases of dispersion law.

### 3. Discussion of results.

Let us proceed to the discussion of results. As it is seen from the above obtained energy expressions, accounting nonparabolicity of the dispersion law, in both 2D and 3D QDs in both SQ regimes leads to a significant change in the energy spectrum of the electron-positron pair in comparison with the parabolic case. Thus, in the case of a semiconductor with a parabolic dispersion (for *GaAs* QD) the dependence of the energy of electron-positron pair on QD sizes is proportional to $\sim \frac{1}{r_0^2}$ ($r_0$ is QD radius), whereas this dependence is violated in the case of Kane dispersion law (for *InSb* QD).

Moreover, in a spherical QD accounting of nonparabolicity of dispersion removes the degeneracy of the energy in the orbital quantum number, and in a circular QD – in the magnetic quantum number. As is known, the degeneracy in the orbital quantum number is a result of the hidden symmetry of the Coulomb problem [44]. From this point of view, the lifting of degeneracy is a consequence of lowering symmetry of the problem, which is in it turn a consequence of the reduction of the symmetry of the dispersion law of the charge carriers, but not a reduction of the geometric symmetry. This results from the narrow-gap semiconductor *InSb* bands interaction. In other words, with the selection of specific materials, for example, *GaAs* or *InSb*, it is possible to decrease the degree of "internal" symmetry of the sample without changing the external shape, which fundamentally changes the physical properties of the structure. Note also that maintaining twofold degeneracy in the magnetic quantum number in cases of both dispersion laws is a consequence of retaining geometric symmetry. On the other hand, accounting of nonparabolicity combined with a decrease in the dimensionality of the sample leads to a stronger expression of the sample "internal" symmetry reduction. Thus, in 2D case the energy of Ps atom with Kane dispersion law becomes imaginary. In other words, 2D Ps atom in *InSb* is unstable – it annihilates immediately.

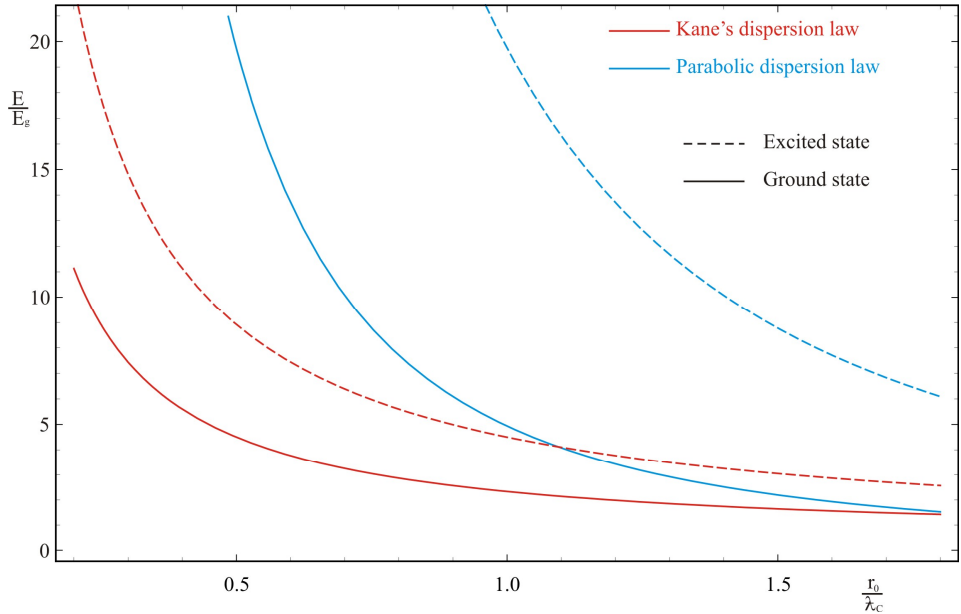

Fig. 2. The dependences of the ground and first excited state energies of the electron-positron pair in a spherical QD on a QD radius in strong SQ regime.

The opposite picture is observed in the case of a parabolic dispersion law. In this case the

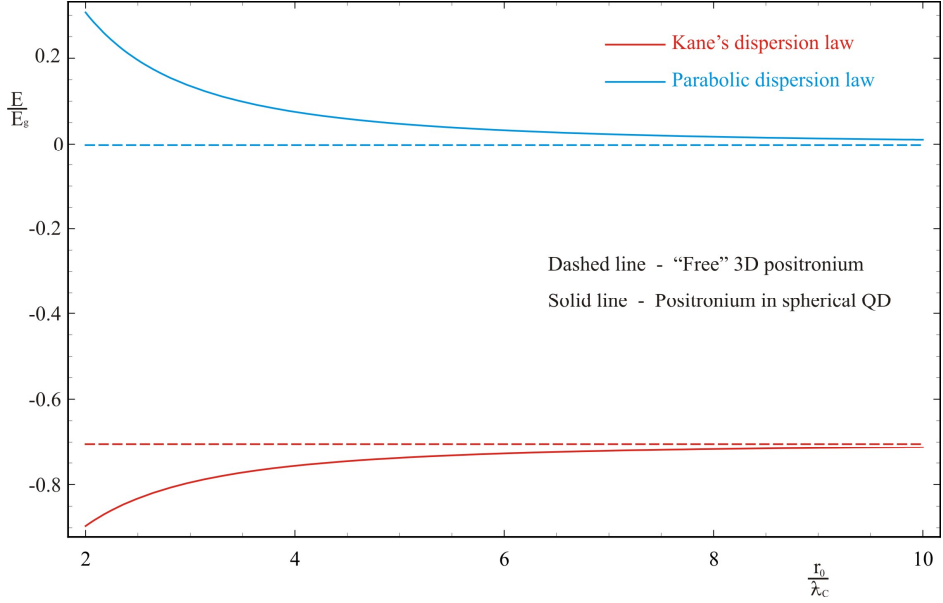

Fig. 3. The dependence of the Ps energy on a QD radius in a spherical QD in the weak SQ regime.

Ps binding energy increases up to 4 times, which in turn should inevitably lead to an increase in a Ps lifetime. It means that it is possible to control the duration of the existence of an electron-positron pair varying the material, dimension and SQ.

Fig. 2 shows the dependences of the ground and first excited state energies of the electron-positron pair in a spherical QD on the QD radius in strong SQ regime. Numerical calculations are made for the QD consisting of *InSb* with the following parameters: $m_e^* = m_p^* \simeq 0.013 m_0$, $E_R^P \simeq 3 \cdot 10^{-4} eV$, $E_g \simeq 0.23 eV$, $\kappa = 17.8$, $a_p \simeq 10^3 \text{Å}$, $\lambdabar_C \simeq 89 \text{Å}$, $\alpha_0 \simeq 0.123$. As it is seen from the figure, at the small values of QD radius the behavior of curves corresponding to the cases of parabolic and Kane dispersion laws significantly differ from each other. The energies of both cases decrease with increase in a QD radius and practically merge as a result of decreasing the SQ influence. The discrepancy of curves appears sharper in smaller values of QD radius, because the dependence on QD sizes is proportional $\sim \frac{1}{r_0^2}$ in parabolic case, and in the Kane dispersion case the analogous dependence appears under the square root (see (10)). The slow growth of the particles energy with decrease in a QD radius in the case of Kane dispersion law is caused exactly by this fact. The situation is similar for excited states of both cases; however the energy difference is considerably strong. Thus, at $R_0 = \lambdabar_C$ the energy difference of ground states of parabolic and Kane dispersion cases is $\Delta E_{ground} \simeq 2.6 E_g$, whereas for excited states it is $\Delta E_{excited} \simeq 15.24 E_g$.

The dependence of the energy of electron-positron coupled pair – a positronium – on a QD radius in a spherical QD in the weak SQ regime is illustrated in the Fig. 3. As it is seen from the figure, in the weak SQ regime, when the Coulomb interaction energy of particles significantly prevails the SQ energy of QD walls, the Ps energy curves behaviors in parabolic and Kane dispersion cases differ radically. With decrease in radius the energy of the Ps changes the sign and becomes positive in the parabolic case (see (28)). This is a consequence of SQ and Coulomb quantization competition. The situation is opposite in the case of the two-band Kane model approximation. In this case, the decrease in the radius alternates the Coulomb quantization due to the band interaction. In other words, in the case of nonparabolic dispersion law the

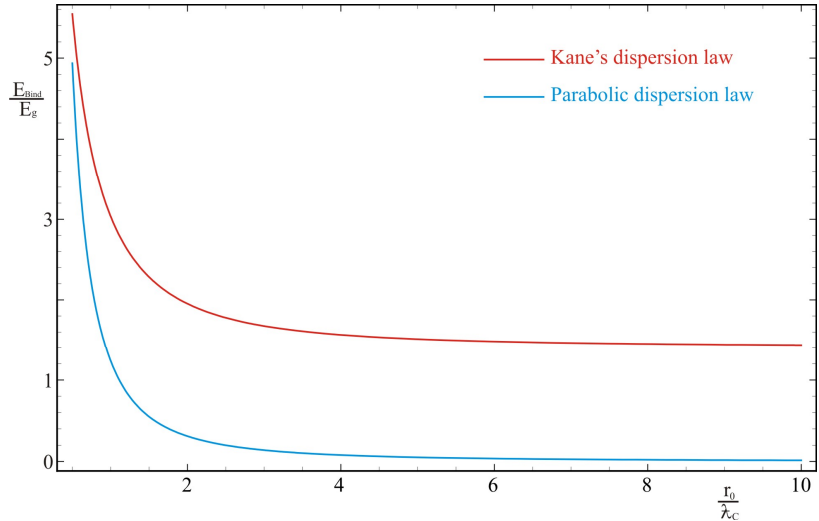

Fig. 4. The dependence of the Ps binding energy in a spherical QD on a QD radius.

Coulomb interaction is stronger [37]. With the increase in the radius both curves tend to the limit of "free" Ps atoms of corresponding cases (these values are given in dashed lines). Sharp increase in the Coulomb interaction in the case of nonparabolicity accounting in the particles dispersion law becomes more apparent from comparison of the dashed lines.

Fig. 4 illustrates the dependence of Ps binding energy in a spherical DQ on the QD radius for both dispersion laws. As it is seen in the figure, with increase in QD radius the binding energy decreases in both cases of dispersion law. However, in the case of Kane dispersion law implementation the energy decrease is slower, and at the limit $R_0 \to \infty$, the binding energy of nonparabolic case remains significantly greater than in parabolic case. Thus, at $R_0 = 3\hbar_C$ in Kane dispersion case the binding energy is $E_{Bind}^{Kane} \simeq 1.675 E_g$, in parabolic case is $E_{Bind}^{Par} \simeq 0.31 E_g$, and at the value $R_0 = 6\hbar_C$ are $E_{Bind}^{Kane} \simeq 1.482 E_g$ and $E_{Bind}^{Par} \simeq 0.036 E_g$, respectively. Note the similar behavior as for the curves of the particle energies and the binding energies in the case of a 2D circular QD.

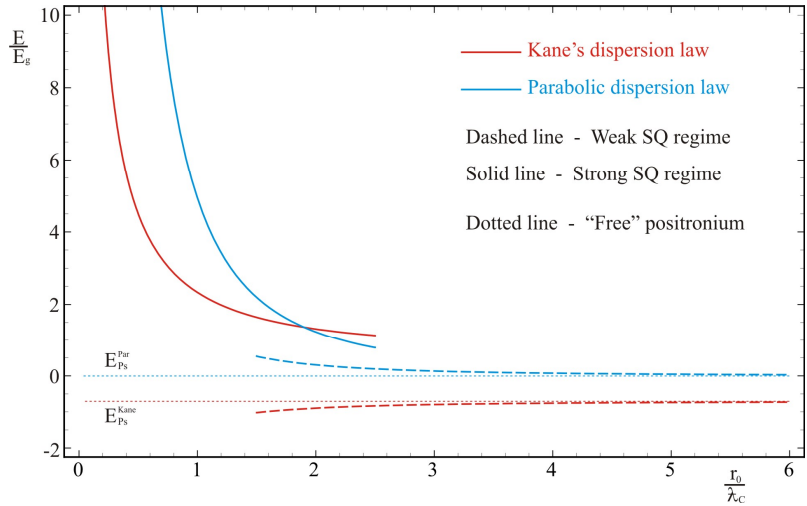

Fig. 5. The dependences of ground state energies on a QD radius for the Ps in weak SQ regime and for separately quantized electron and positron in strong SQ regime.

Finally, Fig. 5 represents the comparative dependences of ground state energies on the QD radius in two QS regimes simultaneously: for Ps in weak SQ regime and for separately quantized electron and positron in strong SQ regime. As shown in the figure, the obtained energy of the coupled electron-positron pair – a positronium – is much smaller than the energy of separately quantized particles. Note that the jump between the energy curves corresponding to strong and weak SQ regimes is precisely conditioned by the formation of Ps atom. This is the criterion of the formation of a Ps as a whole at the particular value of the QD radius. It is seen from the figure that in the case of Kane dispersion law the jump of the energy is significantly greater than in the parabolic case. In other words, more energy is emitted at formation of a Ps in a QD. Consequently, the binding energy of the Ps is much higher than in the case of a parabolic dispersion law. As was noted above, this is a consequence of the Coulomb quantization enhancement due to interaction of bands.

**4. Conclusion.**

In the present paper size-quantized states of the pair of particles – electron and positron – in the strong SQ regime and the atom of Ps in the weak SQ regime were theoretically investigated in a spherical and circular QDs with two-band approximation of Kane dispersion law, as well as with parabolic dispersion law of charge carriers. An additional influence of SQ on Coulomb quantization of a Ps was considered both in 3D and 2D QDs for both dispersion laws. The analytical expressions for the wave functions and energies of the electron-positron pair in the strong SQ regime, and for the Ps as in the weak SQ regime, and in the absence of SQ were obtained in the cases of two dispersion laws and two types of QDs. The fundamental differences between the physical properties of a Ps, as well as separately quantized electron and positron in the case of Kane dispersion law in contrast to the parabolic case were revealed. For the atom of Ps the stability was obtained in a spherical QD and instability in all states with $m=0$ in a circular QD in the case of Kane dispersion law. It was shown that the instability (annihilation) is a consequence of dimensionality reduction, and does not depend on the presence of SQ. More than fourfold increase in the binding energy for the Ps in a circular QD with parabolic dispersion law was revealed compared to the binding energy in a spherical QD. The convergence of the ground state energies and binding energies to the "free" Ps energies for both cases of dispersion laws were shown. It was also revealed the jump between the energy curves corresponding to the cases of strong and weak SQ regimes (which is significantly greater in the case of Kane dispersion law), which is the criterion of the electron and positron coupled state formation – a positronium – at a particular radius of a QD. The removal of a "accidental" Coulomb degeneracy of energy in the orbital quantum number for a spherical *InSb* QD, and in a magnetic quantum number for the circular QD, as a result of a charge carrier dispersion law symmetry degree reduction, was revealed.

**Acknowledgment**

This work is supported by the NSF (HRD-0833184) and NASA (NNX09AV07A).